\documentclass[twocolumn]{el-author}

\newcommand{\ua}{\uparrow}
\newcommand{\nc}{\newcommand}
\nc{\da}{\downarrow} \nc{\hc}{\hat{c}} \nc{\hS}{\hat{S}}
\nc{\bra}{\langle} \nc{\ket}{\rangle} \nc{\eq}{equation (\ref}
\nc{\h}{\hat} \nc{\hT}{\h{T}}\nc{\be}{\begin{eqnarray}}
\nc{\ee}{\end{eqnarray}}\nc{\rd}{\textrm{d}}\nc{\e}{eqnarray}\nc{\hR}{\hat{R}}\nc{\Tr}{\mathrm{Tr}}
\nc{\tS}{\tilde{S}}\nc{\tr}{\mathrm{tr}}\nc{\8}{\infty}\nc{\lgs}{\bra\ua,\phi|}\nc{\rgs}{|\ua,\phi\ket}
\nc{\hU}{\hat{U}}\nc{\lfs}{\bra\phi|}\nc{\rfs}{|\phi\ket}\nc{\hZ}{\hat{Z}}\nc{\hd}{\hat{d}}\nc{\mD}{\mathcal{D}}
\nc{\bd}{\bar{d}}\nc{\bc}{\bar{c}}\nc{\mc}{\mathcal}\nc{\ea}{eqnarray}\nc{\mG}{\mathcal{G}}\nc{\bce}{\begin{center}}
\nc{\ece}{\end{center}}
\date{xxx 2013}

\usepackage[width=18.46cm,height=10in]{geometry}
\usepackage[noadjust]{cite}
\usepackage{srcltx}
\usepackage{psfrag}
\usepackage{epsfig}
\usepackage{amsmath}

\begin{document}

\title{Hybrid-cascade Coupled-Line Phasers for High-resolution Radio-Analog Signal Processing}

\author{ T. Paradis, S. Gupta,  Q. Zhang, L. J. Jiang, and C. Caloz}

\abstract{A hybrid-cascade (HC) coupled-line phaser configuration is presented to synthesize enhanced group delay responses for high-resolution Radio-Analog Signal Processing (R-ASP). Using exact analytical transfer functions, the superiority of HC coupled-line phasers over conventional transversally cascaded C-section phasers is demonstrated and verified using full-wave simulations.}

\maketitle

\section{Introduction}
Radio-Analog Signal Processing (R-ASP) is defined as the real-time manipulation of signals in their analog form to realize specific operations enabling microwave or millimeter-wave and terahertz applications ASP. The heart of an R-ASP system is a \textit{phaser}, which is a component exhibiting a specified frequency-dependent group-delay response within a given frequency range. Such phasers have found useful applications in the domain of instrumentation, security, short-range communication and RADAR systems, to name a few \cite{ASP_MM}.

A common requirement in high-resolution R-ASP is to achieve high dispersion in phasers, in order to provide large frequency discrimination in the time-domain \cite{CRLH_Phaser_TMTT}. This translates into the requirement to achieve a large group delay swing in the phaser within the desired bandwidth. Recently, a coupled-line phaser based on composite right/left-handed (CRLH) transmission lines was proposed utilizing its tight coupling property to achieve this goal \cite{CRLH_Phaser_TMTT}. However, considering its design complexity, multi-layer configuration and fabrication sensitivity, a new configuration based on hybrid-cascaded (HC) coupled-lines was proposed in \cite{Gupta_ICEAA}. This configuration exploits the fact that the required coupling level can be relaxed by folding the conventional C-section into a cross-shaped structure, where superior effective backward coupling results from the discontinuities introduced by the arms of the cross. Consequently, tight coupling coefficients are achieved using different coupled-line sections of relatively small coupling values \cite{TightCoupler_TMTT}. This paper demonstrates the advantage of using HC coupled-line phasers over regular cascaded C-section phasers in group delay engineering via both analytical and full-wave simulation results.

\section{C-section Phasers}

\begin{figure}[h]
\centering
\subfigure[]{
\psfrag{a}[c][c][1]{$(2m+1)\lambda_g/4$@$f_0$}
\psfrag{b}[c][c][1]{$\theta$}
\psfrag{c}[c][c][1]{$k_0$}
\psfrag{z}[c][c][1]{$k$}
\psfrag{x}[c][c][1]{$s_1$}
\psfrag{y}[c][c][1]{$s_2$}
\psfrag{d}[c][c][1]{4-port coupler}
\psfrag{e}[c][c][1]{C-section phaser}
\psfrag{f}[c][c][1]{\shortstack{hybrid-cascade \\coupled-line}}
\includegraphics[width=\columnwidth]{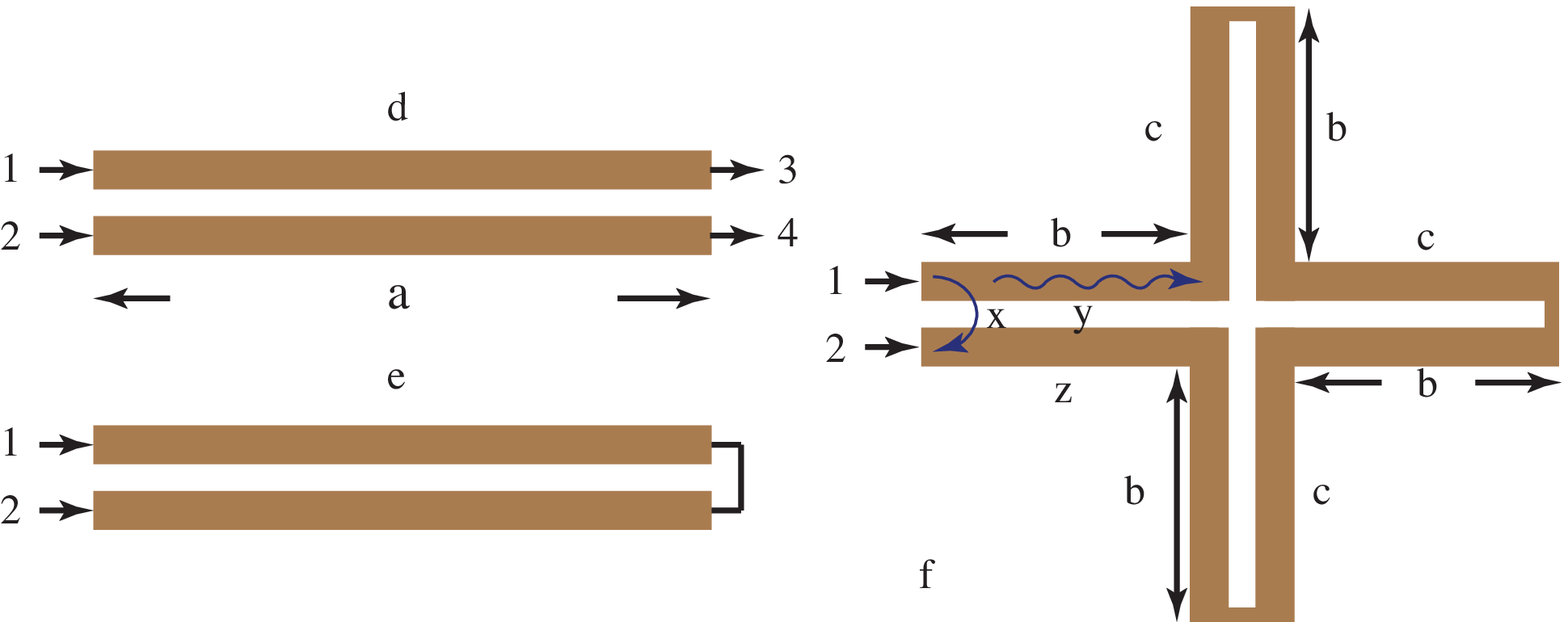}}
\subfigure[]{
\psfrag{a}[c][c][1]{$a$}
\psfrag{b}[c][c][1]{$b$}
\psfrag{s}[c][c][0.8]{$S_0$}
\psfrag{e}[l][c][0.8]{analytical}
\includegraphics[width=\columnwidth]{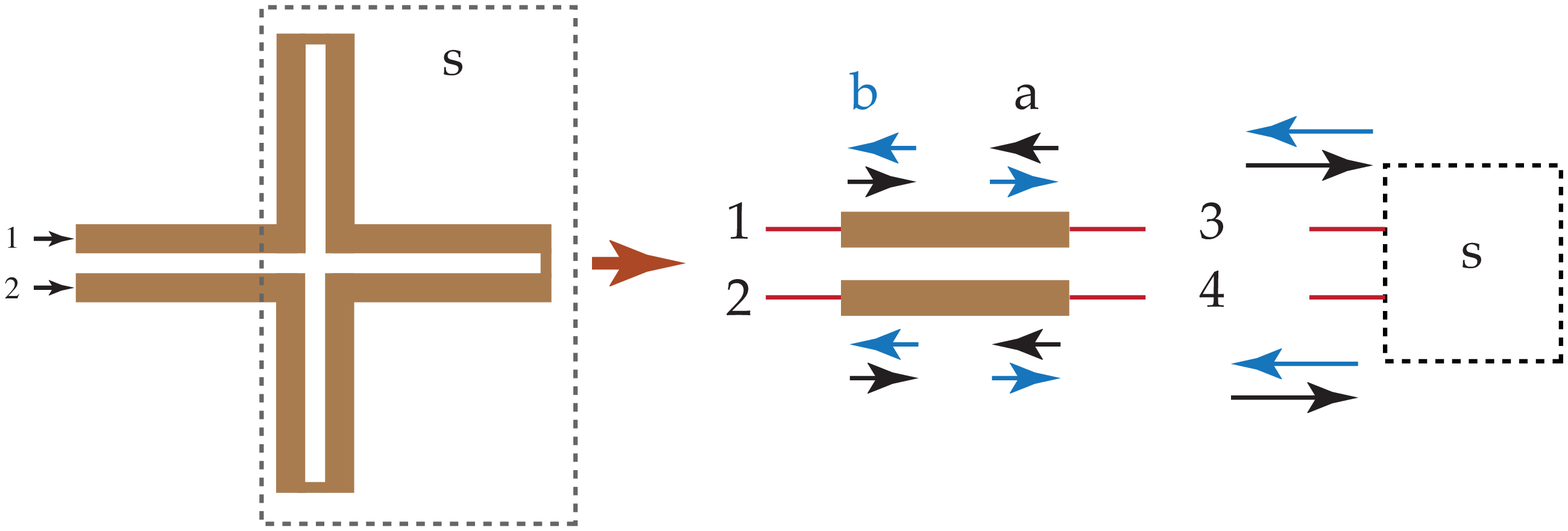}
}
\caption{C-section all-pass phaser configurations. a) C-section phaser formed from a single coupled-line coupler and hybrid-cascade (HC) coupled-line phaser consisting of 3 C-sections and 1 coupled-line coupler. b) HC coupled-line phaser configuration analyzed as a coupled-line coupler terminated with a frequency dependent load $S_0$.}\label{Fig:CSectionPhasers}
\end{figure}

A C-section is a coupled-line coupler with its two ends interconnected so as to form a two-port network, as shown on the left of Fig.~\ref{Fig:CSectionPhasers}(a). It is an all-pass network ($|S_{21}(\omega)|=1\forall\omega$) exhibiting a frequency dependent group delay response, with its delay maxima occurring at odd-multiple of quarter-wavelength frequencies, i.e. $\ell = (2m+1)\lambda_g$ \cite{Cristal_commensurate_synthesis}\cite{Steenaart_CD_ComplexPoles}. Such a conventional C-section can be extended to a higher-order configuration, as proposed in \cite{Gupta_ICEAA}, to form an HC coupled-line phaser, as shown in the right of Fig.~\ref{Fig:CSectionPhasers}(a). This HC coupled-line phaser may be seen as a cascading of a coupled-line coupled loaded with transversally-cascaded C-sections. As shown in \cite{Gupta_ICEAA}, it provides a large group delay swing, as a result of enhanced effective-coupling between the coupled lines, which allows one to use coupled-line sections with moderate couplings.

\begin{figure}[h]
\centering
\psfrag{a}[c][c][1]{frequency, $f$ (GHz)}
\psfrag{b}[c][c][1]{group delay, $\tau(f)$}
\psfrag{c}[l][c][0.8]{Cascaded C-section}
\psfrag{d}[l][c][0.8]{HC coupled-line}
\includegraphics[width=0.7\columnwidth]{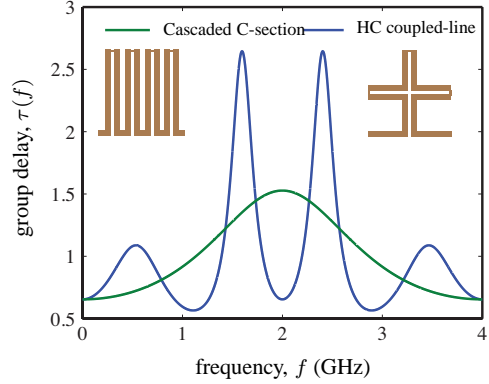}
\caption{Typical group delay responses for a conventional and a HC coupled-line phaser. The unfolded length in each case is $4(2\theta)$ and the identical coupling coefficient $k=0.4$ in all the coupled sections in both phasers.}\label{Fig:Hybrid}
\end{figure}

Assuming that the coupled lines are ideal lossless, perfectly matched and perfectly isolated TEM lines, the 4-port scattering matrix of the input coupled-line coupler of the HC phaser reads \cite{Mongia-Coupler-Book}

\begin{align}
\left[\begin{array}{c}
b_1\\
b_2\\
b_3\\
b_4\\
\end{array}\right]=
\left[\begin{array}{cccc}
0 & s_1(\theta) & s_2(\theta) & 0\\
s_1(\theta) & 0 & 0 & s_2(\theta)\\
s_2(\theta) & 0 & 0 & s_1(\theta)\\
0 & s_2(\theta) & s_1(\theta) & 0\\
\end{array}\right]
\left[\begin{array}{c}
a_1\\
a_2\\
a_3\\
a_4\\
\end{array}\right],\label{Eq:4portCC}
\end{align}

\noindent with

\begin{subequations}
\begin{equation}
s_1(\theta) = \frac{jk\sin\theta}{\sqrt{1-k^2}\cos\theta + j\sin\theta},
\end{equation}
\begin{equation}
s_2(\theta) = \frac{\sqrt{1-k^2}}{\sqrt{1-k^2}\cos\theta + j\sin\theta},
\end{equation}\label{Eq:s_1_s_2}
\end{subequations}

\noindent where $k$ is the voltage coupling coefficient. For the two- port frequency-dependent load $S_0$, represented in the left part in Fig. 1(b), the following relation holds:

\begin{align}
\left[\begin{array}{c}
a_3\\
a_4\\
\end{array}\right]=
\left[\begin{array}{cc}
0 & S_0\\
S_0 & 0\\
\end{array}\right]
\left[\begin{array}{c}
b_3\\
b_4\\
\end{array}\right],\label{Eq:2portCC}
\end{align}

\noindent which transforms the 4-port coupled-line coupler  into a 2-port C-section described, using (\ref{Eq:4portCC}) and (\ref{Eq:2portCC}),  by the following set of equations: $b_1 = s_1a_2 + s_2S_0b_4$, $b_2  = s_1a_1 + s_2S_0b_3$, $b_3 = s_2a_1/(1-S_0s_1)$ and $b_4 = s_2a_2/(1-S_0s_1)$. These equations finally lead to the 2-port transfer function

\begin{equation}
S_{21}(\theta) = \frac{b_2}{a_1} = \frac{b_1}{a_2} = \left(s_1 + \frac{s_2^2S_0}{1-S_0s_1}\right).\label{Eq:S21_S0}
\end{equation}

\noindent When $S_0=1$, a regular C-section is obtained with the transfer function

\begin{align}
S_{21}(\theta) = \left(\frac{\rho - j\tan\theta}{\rho + j\tan\theta}\right)\quad\text{where}\quad \rho = \sqrt{\frac{1 +k }{1 - k}}.\label{Eq:CSection}
\end{align}

\noindent The transfer function of the HC coupled-line coupler of Fig.~\ref{Fig:CSectionPhasers}(a) is given by (\ref{Eq:S21_S0}) with the termination transfer function 

\begin{equation}
S_0 = \left(\frac{\rho_0 - j\tan\theta}{\rho_0 + j\tan\theta}\right)^3\quad \text{where}\quad\rho_0 = \sqrt{\frac{1 +k_0}{1 - k_0}}.
\end{equation}

Figure~\ref{Fig:Hybrid} shows the comparison between the typical group delay response of a conventional C-section and an HC coupled-line phaser, for the case of equal unfolded lengths. As can been seen, the HC coupled-line phaser exhibits a larger group delay swing for the same amount of maximum coupling-coefficient $k$, at the expense of a small bandwidth around each delay peak. Therefore, for low bandwidth applications, HC coupled-line phasers offer a distinct benefit of large delay swing, and thus high dispersion, compared to transversal cascade of C-sections.

\section{Group Delay Engineering Example}

To illustrate the benefit of HC coupled-line phasers, its role in synthesizing a specified group delay response using coupled-line phasers is investigated. Lets consider the objective of synthesizing a linear group delay response with a delay swing of $\Delta\tau = 3$~ns within 2-3 GHz bandwidth, under the constraint that $k < 0.4$. The coupling constraint ensures that an edge-coupled structure is sufficient to realize the final phaser in a planar configuration.

\begin{figure}[h]
\centering
\psfrag{a}[c][c][1]{frequency (GHz)}
\psfrag{b}[c][c][1]{group delay (ns)}
\psfrag{d}[l][c][0.8]{specification}
\psfrag{e}[l][c][0.8]{analytical result}
\includegraphics[width=\columnwidth]{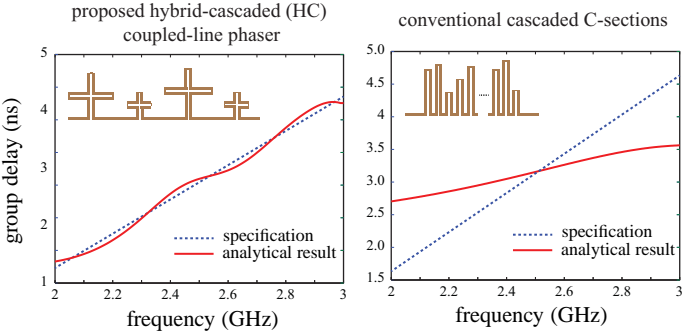}
\caption{Analytical group delay response of a cascaded conventional C-section and HC coupled-line phaser to realize a specified linear-group delay response. Coupling coefficient $k < 0.4$ in all cases. The design parameters for the HC coupled-lines are [see Fig.~\ref{Fig:CSectionPhasers}(a)]: $k_i = (0.25,\;    0.4,\;    0.33,\;    0.4)$, $k_{0,i}=(0.4, \;   0.4,\;    0.4,\;   0.4)$ and $f_{0,i}= (3.05, \;   3.77,  \;  3.44, \;   3.77)$.}\label{Fig:Analytical}
\end{figure}

Fig.~\ref{Fig:Analytical}(a) shows that the specified delay response can be successfully achieved by using a 4-HC coupled-line phaser of different length and coupling coefficients. The previous approach consists in using a transversal cascade of C-sections of different lengths and couplings, as depicted in \cite{Gupta_Phase_Synthesis}. However, as seen from Fig.~\ref{Fig:Analytical}(b), this configuration fails to achieve the objective, the specified delay slope being too large to be realizable using this approach. Therefore, under the constraints of a maximum allowed $k$ and maximum unfolded length, the HC coupled-line phaser is a superior choice.

\begin{figure}[h]
\centering
\psfrag{a}[c][c][1]{frequency (GHz)}
\psfrag{b}[c][c][1]{group delay (ns)}
\psfrag{c}[c][c][1]{S-parameters (dB)}
\psfrag{d}[l][c][0.8]{specification}
\psfrag{e}[l][c][0.8]{MoM ADS}
\psfrag{f}[l][c][0.8]{$S_{21}$}
\psfrag{g}[l][c][0.8]{$S_{11}$}
\psfrag{x}[c][c][0.8]{$n=1$}
\psfrag{y}[c][c][0.8]{$n=2$}
\psfrag{z}[c][c][0.8]{$\ell_0$}
\psfrag{m}[c][c][0.8]{$h = 50$~mil}
\psfrag{n}[c][c][0.8]{$\varepsilon_r = 10.2$}
\psfrag{p}[c][c][0.8]{$\ell_1$}
\psfrag{q}[c][c][0.8]{$\ell_i\in[1\;2 \;3 \;4 ]~mm$}
\psfrag{r}[c][c][0.8]{$g_i\in[1\;2 \; 3\; 4 ]~mm$}
\psfrag{q}[l][c][0.8]{\shortstack{$\ell_i = (7.58,\;	6.17,\; 6.57,\;6.17)$mm\\$g_{i,\text{coupled}}=(0.1,\;0.12,\;0.16,\;0.12)$mm\\$w_{i,\text{coupled}}=(0.16,\;0.14,\;0.15,\;0.14)$mm\\$g_{i,\text{C-section}}=(0.11,\;0.11,\;0.11,\;0.11)$mm\\$w_{i,\text{C-section}}=(0.14,\;0.14,\;0.14,\;0.14)$~mm.}}
\includegraphics[width=\columnwidth]{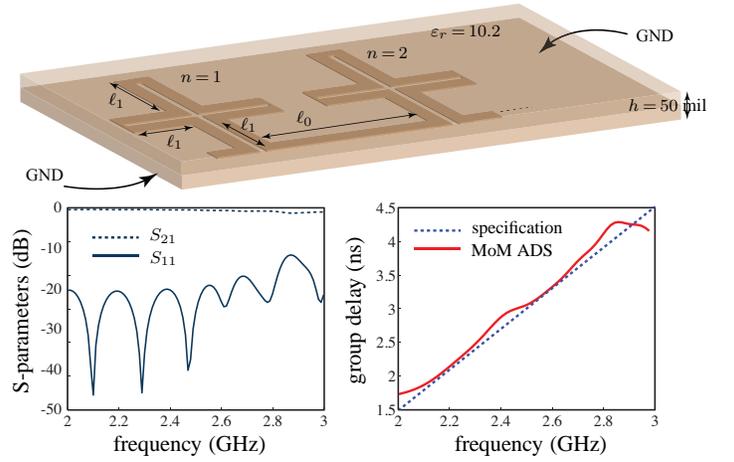}
\caption{Full-wave computed S-parameters and group delay response of a HC coupled-line phaser ($N=4$), implemented in stripline technology, to realize a specified linear-group delay response. Coupling co-efficient $k < 0.4$ in all cases. The physical parameters of the structure are: $\ell_i = (7.58,\;	6.17,\; 6.57,\;6.17)$mm, $g_{i,\text{coupled}}=(0.1,\;0.12,\;0.16,\;0.12)$mm, $w_{i,\text{coupled}}=(0.16,\;0.14,\;0.15,\;0.14)$mm, $g_{i,\text{C-section}}= 0.11$mm $\forall i$ and $w_{i,\text{C-section}}=0.14$~mm $\forall i$.}\label{Fig:FullWave}
\end{figure}

To validate the synthesized delay response using the HC coupled-line phaser, a corresponding stripline circuit was modelled in ADS Momentum, using edge-coupled transmission lines only, as illustrated in Fig.~\ref{Fig:FullWave}. The structure consists of a copper trace sandwiched between two dielectric layers of Rogers RO3010 with $\varepsilon_r = 10.2$ and $25$~mil thickness. A satisfactory return loss with $S_{11} < -10$~dB~$\forall f$ is achieved, along with a successful realization of the specified linear group delay response.

\section{Conclusion}

A hybrid-cascade (HC) coupled-line phaser configuration has been demonstrated for high-resolution R-ASP.  Wheras the HC coupled-line phaser easily achieved the desired response, the conventional C-section phaser failed to realize the required group delay slope. 

\vskip3pt
\ack{This work was supported by NSERC Grant CRDPJ 402801-10 in partnership with Blackberry Inc.}

\vskip5pt

\noindent T. Paradis and C. Caloz (\textit{The Department of Electrical Engineering, PolyGrames Research Center, \'{E}cole Polytechnique de Montr\'{e}al, Montr\'{e}al, QC, Canada.})
\vskip3pt
\noindent Q. Zhang (\textit{Electrical Engineering Department of South University of Science and Technology of China, Shenzhen, China.})
\vskip3pt
\noindent S. Gupta and L. J. Jiang (\textit{Department of Electrical and Electronic Engineering, The University of Hong Kong, Hong Kong,  China.})
\vskip3pt

\noindent E-mail: shulabh@hku.hk

\bibliographystyle{aer}
\bibliography{Paradis_HybridPhaser_EL_2013_Ref}

%
%
%
%

\end{document}